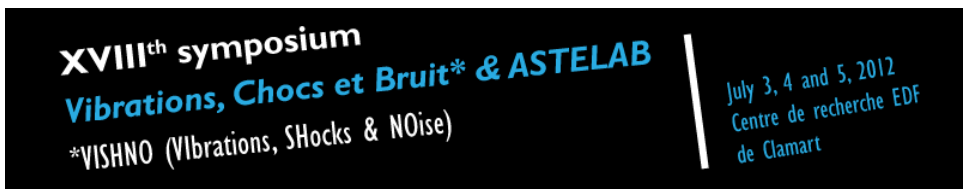

# Nonlinear normal modes of a two degree of freedom oscillator with a bilateral elastic stop

E. H. Moussi[a,b,]*, S. Bellizzi[a], B. Cochelin[a], I. Nistor[b]

[a] *Laboratoire de Mécanique et d'Acoustique (LMA), CNRS, UPR 7051, Aix-Marseille Univ, Centrale Marseille, F-13402 Marseille Cedex 20, France*
[b] *Laboratoire de Mécanique des Structures Industrielles Durables (LaMSID), UMR EDF-CNRS-CEA 2832, 1 Avenue du Général de Gaulle, 92141 Clamart, France*

**Abstract**

A study of the non linear modes of a two degree of freedom mechanical system with bilateral elastic stop is considered. The issue related to the non-smoothness of the impact force is handled through a regularization technique. In order to obtain the Nonlinear Normal Mode (NNM), the harmonic balance method with a large number of harmonics, combined with the asymptotic numerical method, is used to solve the regularized problem. These methods are present in the software "package" MANLAB. The results are validated from periodic orbits obtained analytically in the time domain by direct integration of the non regular problem. The two NNMs starting respectively from the two linear normal modes of the associated underlying linear system are discussed. The energy-frequency plot is used to present a global vision of the behavior of the modes. The dynamics of the modes are also analyzed comparing each periodic orbits and modal lines. The first NNM shows an elaborate dynamics with the occurrence of multiple impacts per period. On the other hand, the second NNM presents a more simple dynamics with a localization of the displacement on the first mass.

*Keywords: nonlinear normal modes, localized impact, internal resonance* ;

## 1. Introduction and industrial issue

In the components of the nuclear power plants facilities, vibrating structures with localized contact conditions are frequently encountered. For instance, in pressurized water reactors, steam generators in which heat is exchanged between the primary and the secondary coolant fluids, that type of contact conditions are usual. In fact, steam generators consist of a bundle of U-tubes, in which flows the primary coolant fluid, and a number of support plates, to guide the tubes. The secondary fluid flows between the tubes and plates, and consequently fill the space with sludge deposits, and then accelerates and increases the excitation of the tubes. This phenomenon can cause dynamical instabilities leading to tube cracks. The linear vibration theory is not sufficient for explaining these phenomena. The use of Nonlinear Normal Modes (NNMs) [1] seems to be a useful framework for investigating the complexity of nonlinear dynamics.

In this context and as a first step of a more general study, we propose here to study a two degree of freedom oscillator with a bilateral elastic stop. This system comes from an analogy with a simplified model of a U-tube as shown in Fig. 1 where the elastic stop stiffness is modeled as a piecewise linear law. This system can be considered as a part of the class of the vibro-impact systems. The vibro-impact systems have been thoroughly studied in the literature. The forced response case has often been considered with soft [2] and rigid [3] impacts. The NNMs of a two DOF system with one elastic stop was studied in [4]. The same model was used in [5] but considering rigid and soft impacts and computing free and forced responses. Vibro-impact absorbers have been characterized in terms of the NNMs in [6]. Our study corresponds to an extension of these works.

*El Hadi Moussi. Tel.: +33147653207; fax: +0-000-000-0000 .
*E-mail address*: el-hadi.moussi@edf.fr.

A regularization of the non-smooth system is introduced which permits us to use the nonlinear normal mode theory as defined by Shaw and Pierre [7], and particularly its variants considered in [8] where a family of periodic orbits is sufficient to describe a nonlinear normal mode. To compute the periodic orbits of the system, the Harmonic Balance Method (HBM) combined to the Asymptotic Numerical Method (ANM), as implemented in the "package software" MANLAB, is used.

The paper is organized as follows. In the next section, the non smooth and regularized models are described. Section 3 is dedicated to the computation of the period orbits. In case of the non smooth model, a method is proposed to compute the periodic orbits with two impacts per period. These solutions are then used to validate the regularized model. In case of the regularized system, the HBM and ANM methods are described. Finally, in Section 4, the results for the NNMs are discussed in detail.

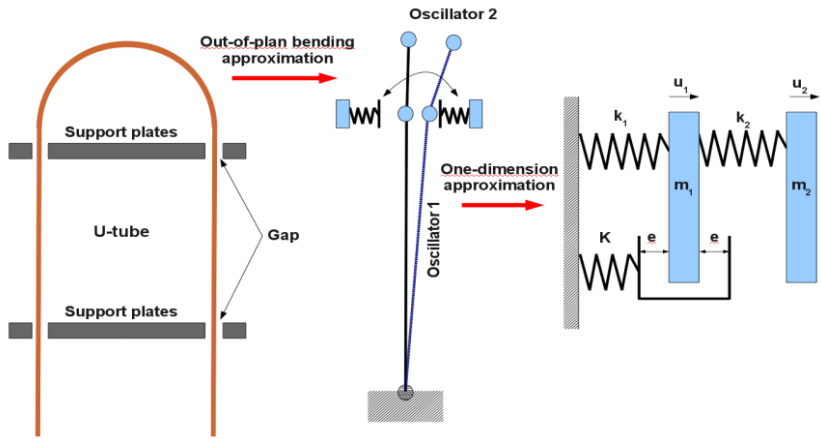

Fig 1. *Analogy of a simplified model of a U-tube with a two degree of freedom oscillator*

## 2. Problem formulation

### 2.1. A two degree of freedom oscillator with an elastic stop

The system under consideration is shown Fig. 1 (right). It consists of two masses $m_1$ and $m_2$ connected by two linear springs of stiffness $k_1$ and $k_2$. The motion of the mass $m_1$ is limited by two identical linear elastic stops with a linear spring of stiffness $K$ and a gap $e$. The equations of motion are given by

$$\begin{cases} m_1\ddot{U}_1(t) + k_1 U_1(t) + k_2\left(U_1(t) - U_2(t)\right) + F\left(U_1(t)\right) &= 0 \\ m_2\ddot{U}_2(t) + k_2\left(U_2(t) - U_1(t)\right) &= 0 \end{cases} \tag{1}$$

with

$$F(U) = \begin{cases} K(U - e) & \text{if} & e \leq U \\ 0 & \text{if} & -e \leq U \leq e \\ K(U + e) & \text{if} & U \leq -e \end{cases} \tag{2}$$

where $U_i$ denotes the displacement of the mass $m_i$ (for $i = 1, 2$) and $F$ denotes the bilateral contact force.

Using now the following rescaled quantities $x = \dfrac{U_1}{e}$, $u = \dfrac{U_2}{e}$, $\hat{f} = \dfrac{F}{k_1 e}$ and the time normalization

$\tau = \omega t$ with $\omega = \sqrt{k_1 m_1^{-1}}$, Eqs. (1)(2) take the following nondimensional form

$$\begin{cases} \ddot{x}(\tau) + x(\tau) + \beta\big(x(\tau) - u(\tau)\big) + \hat{f}\big(x(\tau)\big) & = & 0 \\ \delta\ddot{u}(\tau) + \beta\big(u(\tau) - x(\tau)\big) & = & 0 \end{cases} \tag{3}$$

with

$$\hat{f}(x) = \begin{cases} \alpha(x-1) & \text{if} & 1 \le x \\ 0 & \text{if} & -1 \le x \le 1 \\ \alpha(x+1) & \text{if} & x \le -1 \end{cases} \tag{4}$$

where $\alpha = \dfrac{K}{k_1}$, $\beta = \dfrac{k_2}{k_1}$, $\delta = \dfrac{m_2}{m_1}$ and (.) denotes now the time derivative with respect to the new time $\tau$.

Eqs. (3)(4) only depend on three parameters. The first two parameters characterize the ratios of stiffness and the last one is equal to the ratio of masses. In application, $\alpha$ may be very large for a good representation of the stop, whereas $\beta$ and $\delta$ could be around $1$. Note that the rescaled elastic stop is associated to a gap value equal to $1$. In the sequel, we will restrict the discussion to Eqs. (3)(4).

Note that the set of Eqs. (3)(4) constitutes a piecewise linear system. This remark will be used to build the periodic orbits.

### 2.2. A two degree of freedom oscillator with regularized elastic stop

Regularized equations of motion can be derived approximating the piecewise linear function Eq. (4) by the following polynomial (with respect to the variable $f$)

$$f(f - \alpha(x-1))(f + \alpha(x+1)) = -\eta\alpha^2 x . \tag{5}$$

The positive real $\eta$ denotes the regularization parameter. For $\eta = 0$, the values of $\hat{f}(x)$ appear to be roots of the polynomial Eq. (5). For a given set of parameter values $(x, \alpha, \eta)$, this polynomial always admits a real root denoted $f(x; \alpha, \eta)$ (the expression is not given here) which satisfies the following properties:

(i) $x$ and $f(x; \alpha, \eta)$ have the same sign $\hspace{4em}$ (6)

(ii) for $|x| \ll 1$, $f(x; \alpha, \eta) \approx \eta x$

(iii) for $|x| \gg 1$, $f(x; \alpha, \eta) \approx \alpha(x + \text{sgn}(x))$.

The regularization parameter introduces linear spring of stiffness $\eta$ at the neighbour of the equilibrium point

Finally, the regularized equations of motion defined by

$$\begin{cases} \ddot{x}(\tau) + x(\tau) + \beta\big(x(\tau) - u(\tau)\big) + f\big(x(\tau); \alpha, \eta\big) & = & 0 \\ \delta\ddot{u}(\tau) + \beta\big(u(\tau) - x(\tau)\big) & = & 0 \end{cases} \tag{7}$$

can advantageously replace Eqs. (3)(4) when $\eta$ is small compared to spring stiffness of the underlying linear system.

## 3. Computation of periodic orbits of the system

Two approaches we will use to computed the periodic orbits. One is based on the piecewise linear structure of Eqs. (3)(4) and gives access to the exact synchronous periodic solutions. The other is applied to Eq. (7) and combines the Harmonic Balance Method (HBM) and the Asymptotic Numerical Method (ANM) and gives access to the synchronous and no synchronous periodic solutions.

### 3.1. Two degree of freedom oscillator with an elastic stop case

The definition of the NNM proposed by Rosenberg [Rosenberg, 1966,8] is used here. A NNM is a family of synchronous oscillations (i.e. periodic vibrations in unison: all material points of the system reach their extreme values and pass through zero simultaneously).

We limit the discussion to the case of periodic orbits with two impacts per period (one impact per stop). Starting from the equilibrium point $(x_0, u_0) = (0,0)$, a periodic orbit in the configuration space can be decomposed in four branches:

for $0 \leq t \leq T_1$ where $T_1$ corresponds to $\dot{x}(T_1) = 0$, $\dot{u}(T_1) = 0$ and $x(T_1) = \max_{0 \leq t \leq T_1} x(t)$ with

$x(T_1) > 1$ (extreme values);

- for $T_1 \leq t \leq T_1 + T_2$ where $T_2$ corresponds to $x(T_1 + T_2) = 0$, $u(T_1 + T_2) = 0$;
- for $T_1 + T_2 \leq t \leq T_1 + T_2 + T_3$ where $T_3$ corresponds to $\dot{x}(T_3) = 0, \dot{u}(T_3) = 0$ and

  $x(T_3) = \min_{T_1 + T_2 \leq t \leq T_1 + T_2 + T_3} x(t)$ with $x(T_3) < -1$ (extreme values);

- for $T_1 + T_2 + T_3 \leq t \leq T_1 + T_2 + T_3 + T_4$ where $T_4$ corresponds to $x(T_1 + T_2 + T_3 + T_4) = 0$,

  $u(T_1 + T_2 + T_3 + T_4) = 0$.

Due to the symmetry of the system, it can be established that the branches satisfy

- for $T_1 \leq t \leq T_1 + T_2$, $x(t) = x(2T_1 - t)$ and $u(t) = u(2T_1 - t)$;
- for $T_1 + T_2 \leq t \leq T_1 + T_2 + T_3$, $x(t) = -x(2T_2 - t)$ and $u(t) = -u(2T_2 - t)$;
- for $T_1 + T_2 + T_3 \leq t \leq T_1 + T_2 + T_3 + T_4$, $x(t) = x(2T_3 - t)$ and $u(t) = u(2T_3 - t)$.

It can then be deduced that $T_1 = T_2 = T_3 = T_4$ given the period as $T = 4T_1$ and showing that the period orbit is only characterized by the first branch i.e. only on a quarter period.

Re-writing Eqs. (3)(4) as

-for $-1 \leq x(\tau) \leq 1$,

$$\begin{cases} \ddot{x}(\tau) + x(\tau) + \beta\big(x(\tau) - u(\tau)\big) & = & 0 \\ \delta\ddot{u}(\tau) + \beta\big(u(\tau) - x(\tau)\big) & = & 0 \end{cases}, \tag{8}$$

-for $x(\tau) \geq 1$,

$$\begin{cases} \ddot{x}(\tau) + x(\tau) + \beta\big(x(\tau) - u(\tau)\big) + \alpha x(\tau) & = & \alpha \\ \delta\ddot{u}(\tau) + \beta\big(u(\tau) - x(\tau)\big) & = & 0 \end{cases}, \tag{9}$$

(the case $x(\tau) \leq -1$ is not considered here), the quarter period branch is defined in two steps:

for $0 \leq t \leq \tau_1$, the branch is defined by Eq. (8) with the initial condition $(0, 0, \dot{x}_0, \dot{u}_0)$ and the time duration $\tau_1$ is defined by $x(\tau_1) = 1$;

for $\tau_1 \leq t \leq T_1$ the branch is defined Eq. (9) with the initial condition $(1, 0, \dot{x}(\tau_1), \dot{u}(\tau_1))$ and the time duration $T_1$ is defined by $\dot{x}(T_1) = 0$, $\dot{u}(T_1) = 0$.

Four unknowns $(\dot{x}_0, \dot{u}_0, \tau_1, T_1)$ are needed to characterize the quarter period branch and the unknowns satisfy three equations: $x(\tau_1) = 1$, $\dot{x}(T_1) = 0$ and $\dot{u}(T_1) = 0$. Following [5], analytic expressions of the Cauchy problems associated to Eq. (8) and Eq. (9) which define the non linear equations can be obtained. The expressions are not reported here. The resulting equations can be re-written in terms of a non linear algebraic system and they can be solved using the continuation Asymptotic Numerical Method (ANM)[11].

This procedure described here could be generalized to characterize periodic orbits with more than two impacts per period.

### 3.2. Two degree of freedom oscillator with regularized elastic stop case

The definition of the NNM retained by [8, 9] is used here. For conservative systems, a NNM is a family of periodic orbits. The Harmonic Balance Method (HBM) is used to approximate the period orbits (method often used to calculate periodic responses) and it is combined with the continuation Asymptotic Numerical Method. The ANM gives access to the branches of solution in the form of high order power series. The ANM is highly robust, even when the branch contains sharp turns, which is quite usual during the computation of NNM.

To apply ANM method, it is first necessary to recast the equations of motion into a dynamical system where the nonlinearities are polynomials and at most quadratics [12]. Here the nonlinearity as defined by Eq. (5) is a third order polynomial. It can be rewritten in the quadratic form as

$$\begin{cases} f - \eta\, x - fz = 0 \\ z - \left( \dfrac{f}{\alpha} - x \right)^2 = 0 \end{cases} \tag{11}$$

giving the following quadratic dynamical system

$$\begin{cases} \ddot{x}(\tau) + x(\tau) + \beta\big(x(\tau) - u(\tau)\big) + f(\tau) = 0 \\ \delta\,\ddot{u}(\tau) + \beta\big(u(\tau) - x(\tau)\big) = 0 \\ f(\tau) - \eta\, x(\tau) - f(\tau)z(\tau) = 0 \\ z(\tau) - \left( \dfrac{f(\tau)}{\alpha} - x(\tau) \right)^2 = 0 \end{cases} \tag{12}$$

Next, the displacements and in our case the contact force $f(\tau)$ and the internal variable $z(\tau)$ are written as a truncated Fourier series

$$\begin{cases} u(\tau) = u_0 + \displaystyle\sum_{k=1}^{H} u_c^k \cos(k\omega\tau) + u_s^k \sin(k\omega\tau) \\ x(\tau) = x_0 + \displaystyle\sum_{k=1}^{H} x_c^k \cos(k\omega\tau) + x_s^k \sin(k\omega\tau) \\ f(\tau) = f_0 + \displaystyle\sum_{k=1}^{H} f_c^k \cos(k\omega\tau) + f_s^k \sin(k\omega\tau) \\ z(\tau) = z_0 + \displaystyle\sum_{k=1}^{H} z_c^k \cos(k\omega\tau) + z_s^k \sin(k\omega\tau) \end{cases} \tag{13}$$

where $H$ denotes the number of harmonics and, as classical, $\omega$ denotes the frequency (which is related to the period as $T = 2\pi / \omega$).

Substituting Eq. (13) into Eq. (7) and balancing the 2H+1 terms, one obtains a non-linear algebraic system of equations which can be written under the form $S(U) = L(U) + Q(U, U) = 0$ where $U = (u_0, x_0, f_0, z_0, ..., u_s^H, x_s^H, f_s^H, z_s^H, \omega) \in \mathfrak{R}^{4(2H+1)+1}$, $L$ is linear and $Q$ is quadratic. Finally, the ANM as described in [10] is applied to solve $S(U) = 0$. This method is based on power series expansions of the unknowns $U(a)$ with respect to the path parameter $a$ and it consists in generating a succession of branches, instead of a sequence of points to compute the NNM.

## 4. Study of periodic orbits of the system

The following numerical values: $\beta = 1$, $\delta = 1$ and $\alpha = 30$ have been used to define the non-smooth system. The regularized system has been defined with $\eta = 0.1$ and the HBM has been carried out with $H = 100$ harmonics.

### 4.1. Comparison between the exact and the approximate solution

The objective of this section is to compare the NNMs of the non-smooth (Eqs. (3)(4)) and regularized (Eq. (7)) systems. We focus on the NNM starting at low energy level from the first Linear Normal Mode (LNM) of the underlying linear system corresponding to the resonance frequency (in Hertz) $\omega = 0.618 /(2\pi) = 0.098$.

The NNMs associated to the two systems are compared in Fig. 2 in terms of Frequency Energy Plot (FEP). The energy range considered here corresponds to periodic orbits with zero or one impact on each stop per period. At low energy level, the difference between the two curves is only due to the regularization parameter $\eta$. For high energy level, the curves coincide. Each point of the FEP corresponds to a periodic orbit. In Fig. 3, the period orbits characterizing the NNM are compared for an intermediate energy level (points a) in Fig. 2) and a high energy level (points b) in Fig. 2).

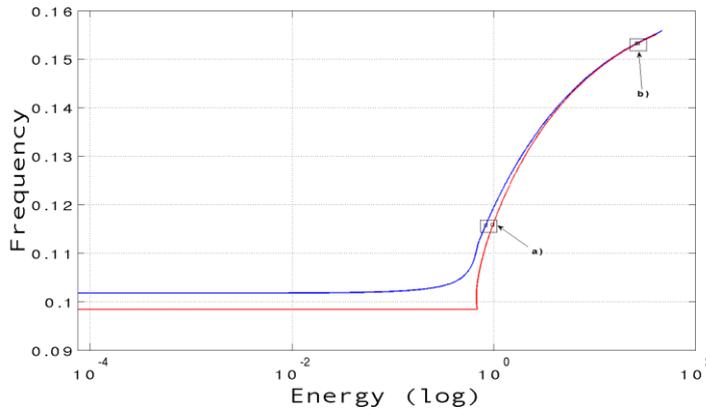

Fig 2. *FEP corresponding to the first NNM of the non-smooth system (in red) and to the regularized one (in blue). (Parameter values:* $\alpha = 30$, $\beta = 1$, $\delta = 1$, $\eta = 0.1$, $H = 100$ *).*

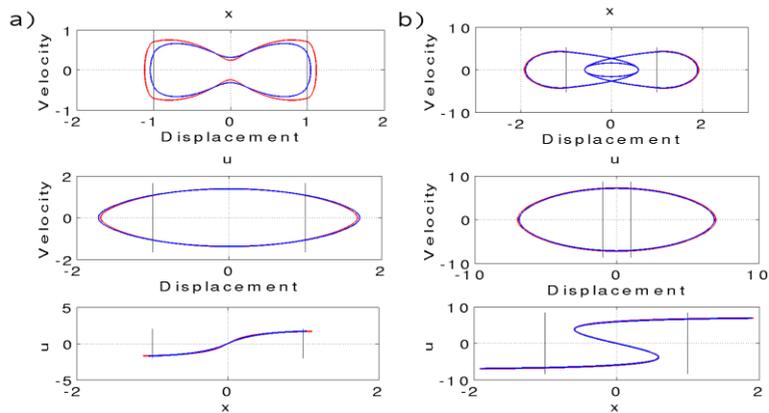

Fig 3. *Periodic orbits at intermediate (a)) and high (b)) energy levels in terms of* $x$ *-phase space (first row),* $u$ *-phase space (second row) and modal line (third row) for the non-smooth system (in red) and the regularized one (in blue). (Parameter values:* $\alpha = 30$, $\beta = 1$, $\delta = 1$, $\eta = 0.1$, $H = 100$ *).*

The regularized solution differs from the piecewise solution for intermediate energy level (see Fig. 3 a)), or equivalently near the natural frequency of the underlying linear system. This result is expected because of the use of the parameter value $\eta = 0.1$. For high energy level (relatively high frequency) (see Fig. 3 b)), the regularized solution is in good agreement with the piecewise solution obtained $H = 100$ harmonics. The behaviour of the contact force is shown Fig. 4. In both cases (intermediate and high energy level), the HBM approximation gives a good approximation.

These results show that the smoothness can be explicitly controlled and seems to give a good approximation even with $\eta = 0.1$.

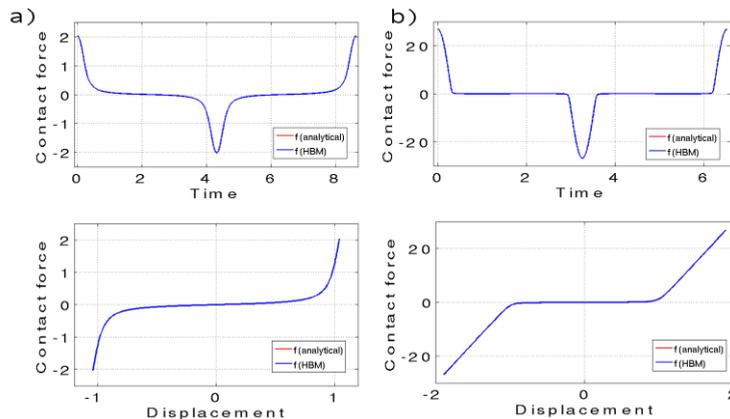

Fig 4. *Contact force at intermediate (a)) and high (b)) energy levels in terms of time evolution over one period (first row) and* $(x, f(x))$ *-plot for the regularized system (in blue) and given by Eq. (6) (in red). (Parameter values:* $\alpha = 30$, $\beta = 1$, $\delta = 1$, $\eta = 0.1$, $H = 100$ *).*

## 4.2. Study of the nonlinear normal modes based on the regularized system

The objective of this section is to analyse the behaviour of the first two NNMs of the regularized system obtained with the method described Section 3.2.

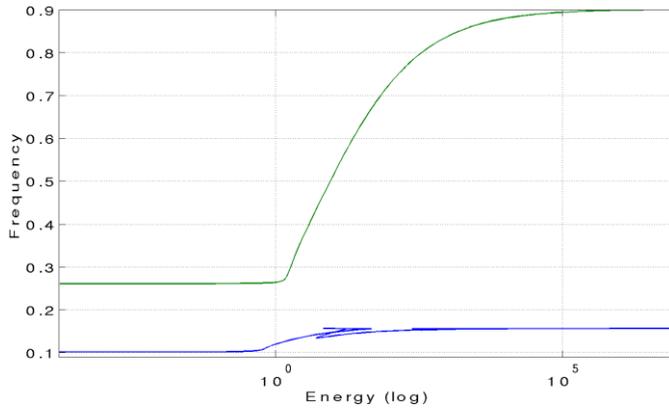

Fig 5. *FEP of the first (in blue) and second (in green) NNM. (Parameter values:* $\alpha = 30$, $\beta = 1$, $\delta = 1$, $\eta = 0.1$, $H = 100$ *)*.

The first two NNM have shown in Fig. 5 in terms of FEP. The first (respectively second) NNM coincides at low energy level with the first (respectively second) resonance frequency of the underlying linear system. Note that due the $n$. the underlying linear system differs from the underlying linear system associated to the non-smooth system ($e = +\infty$). The energy range considered here is larger than the range used in Section 4.1 and periodic orbits with more than one impact on each stop per period are now accessible. The frequency ranges of the two NNMs are separated. The second NNM seems to be simpler in its dynamic. For this reason, it is more convenient to study it, and also because the first NNM has interaction with the second NNM, as we will see later.

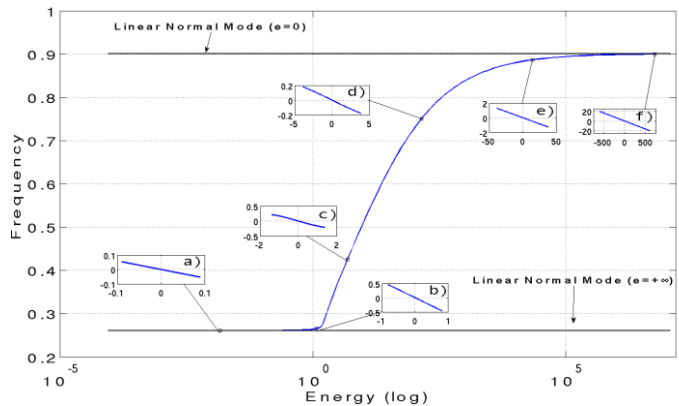

Fig 6. *FEP of the second NNM. (Parameter values:* $\alpha = 30$, $\beta = 1$, $\delta = 1$, $\eta = 0.1$, $H = 100$ *)*.

Fig 6. shows the second NNM where the NNM motions depicted in the configuration space ( modal line) are also reported. At low energy level, the second NNM is constant. The associated frequency is equal to $\omega = 1.639\ /(2\pi) = 0.2611$ and the motions are close to the underlying linear dynamic (see Fig 6. a) and b)), indeed the modal line is a straight line. At high energy level, the curve becomes asymptotic to $\omega = 5.666\ /(2\pi) = 0.9022$ which corresponds to higher resonance frequency of the linear system assuming the gap $e = 0$ and the motions are also close to the associated linear dynamic (see Fig 6. e) and f)). In all the cases, the motions are out of phase. One nonlinear effect observed here is a phenomenon of localization of the motion on the first mass validated by the diminution of the amplitude of the second mass (see Fig 6. c) and d)). Another nonlinear effect can also be observed at intermediate energy level, where the modal lines are now curved.

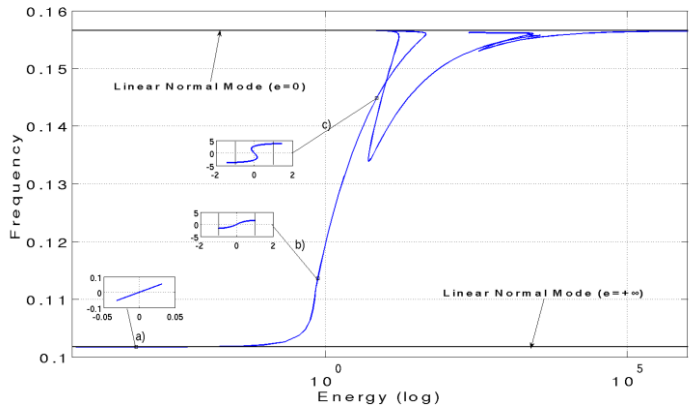

Fig 7. *FEP of the first NNM. (Parameter values: $\alpha = 30$, $\beta = 1$, $\delta = 1$, $\eta = 0.1$, $H = 100$ ).*

The first NNM has a more complicated dynamic. It will be discussed in reference to four energy ranges which are reported successively in Figs 7, 8, 9, 10 and 11 where the NNM motions depicted in the configuration space (modal line) are also shown.

Fig. 7 reproduces the results discussed in Section 4.1. At low energy level, the displacement is close to linear dynamic (see Fig 7. a)) with zero impact on the stops. When the energy increases, one impact on each stop appears and the mode line slightly differs from a straight line (see Fig. 7 b)). Increasing more the energy, a secondary oscillation between each impact is observed (see Fig 7. c)). This oscillation increases in amplitude with the energy level and reaches the two stops (see Fig 8.d)) defining a motion with three impacts on each stop per period. Note that now the frequency is close to the smaller resonance frequency of the linear system assuming the gap $e = 0$.

Fig. 8 details the branch corresponding to periodic orbits with three and two impacts on each stop per period. From the point d), following the branch the energy level decreases up to the point e) where a bifurcation due to a 3:1 internal resonance between the first and second NNMs is observed (see Fig. 8).

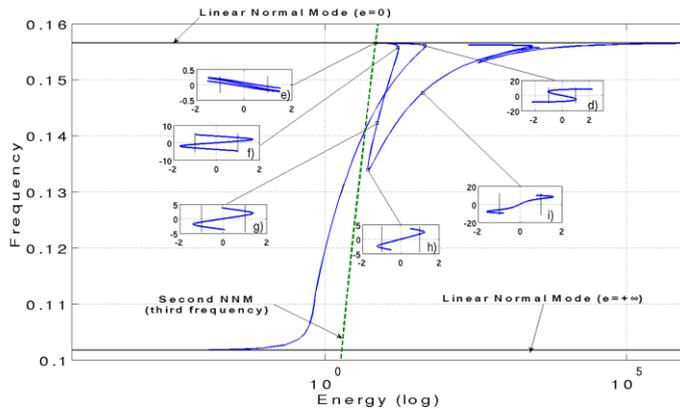

Fig 8. *FEP of the first NNM. The green line corresponds to three time FEP of the second NNM. (Parameter values: $\alpha = 30$, $\beta = 1$, $\delta = 1$, $\eta = 0.1$, $H = 100$ ).*

This bifurcation involve an out-of-phase of the motion (see Fig 8.d) and Fig 8. f)). Before the bifurcation point, the amplitude of the displacement of the second mass decreases. After the bifurcation, the amplitude of the displacement of the second mass increases. At the point f), the number of impact is reduced with from now two impacts on each stop per period. This change of dynamic involved again a diminution of the amplitude of the second mass (Fig 8. f) and g)). From the point h), the scenario observed in Fig. 7 will be reproduced again. Increasing more the energy, a secondary oscillation between each impact is observed (see Fig 8. i) and Fig. 9. j)). This oscillation increases in amplitude with the energy level and reaches the two stops (see Fig 8.d)) defining a motion with four impacts on each stop per period. The difference, here, is that an even number of impacts is reached. So the bifurcation observed before cannot occur, in fact due to the excessive symmetry of the system only the odd harmonic are present.

Figs. 9 and 10 detail the branch corresponding to periodic orbits with four and five impacts on each stop per period. This absence of bifurcation permitted a continuation of increase of the amplitude, and the number of impact reached five impacts per period (see Fig. 9 k)). A bifurcation due to a 5:1 internal resonance between the first and second NNMs is observed (see Fig. 9)). The change to an out-of-phase motion occurred again (see Fig. 9. k) and m)).

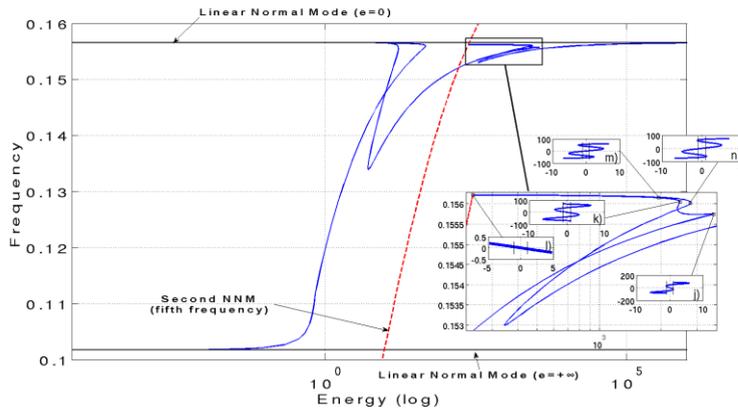

Fig 9. *FEP of the first NNM. The red line corresponds to five time FEP of the second NNM. (Parameter values:* $\alpha = 30$, $\beta = 1$, $\delta = 1$, $\eta = 0.1$, $H = 100$ *).*

The same kind of phenomenon observed before (see Fig. 8 g), h) and i)) can be observed here (see Fig. 10 o), p) and q)). A difference here is the presence of a sequence of impacts without oscillation between the two stops. The phenomenon described below involves a diminution of the number of impact to one on each stop per period.

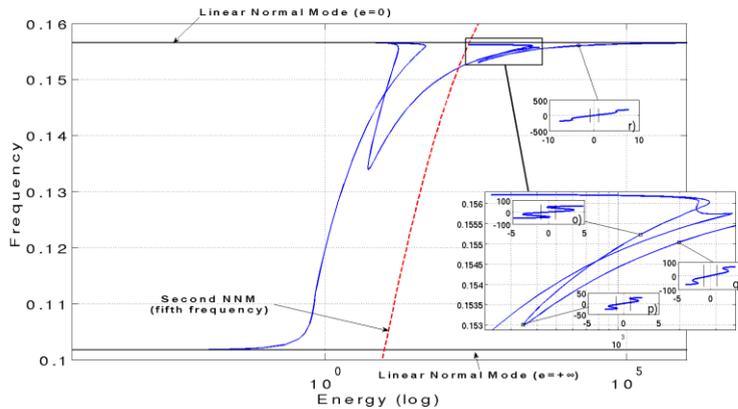

Fig 10. *FEP of the first NNM. The red line corresponds to five time FEP of the second NNM. (Parameter values:* $\alpha = 30$, $\beta = 1$, $\delta = 1$, $\eta = 0.1$, $H = 100$ *).*

As shown Fig.11 when the energy level increases, the dynamic becomes close to the LNM of the linear system with $e = 0$.

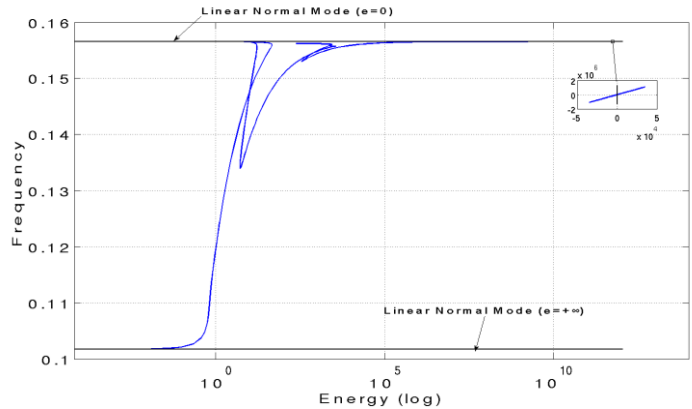

Fig 11. *FEP of the first NNM. (Parameter values:* $\alpha = 30$, $\beta = 1$, $\delta = 1$, $\eta = 0.1$, $H = 100$ *).*

## 5. Conclusion

A study of a two degree of freedom oscillator with bilateral elastic stop corresponding to a simplified model of an industrial structure was presented on this paper. The aim of this work was to compute and analyze the behavior of the NNMs. The choice has been made to use the combination of the HBM and the ANM methods for the computation of the NNMs. To overcome the issue of the non-smoothness of the contact force, a regularization procedure was applied. The influence of the regularization parameters was studied using the computation of the piecewise linear system using only the periodic orbits with one impact on each stop per period. This study has validated the use of this kind of numerical procedures for computation of the NNMs in cases involving a complex dynamics with multiple impacts and internal resonance. The next step is to propose a study of the stability of the NNMs, and to apply this tool to continuous structures discretized by the Finite Element Method.